\documentclass[12pt, a4paper]{article}

\usepackage{amssymb, amsmath}  %%%%%my addition
\usepackage{fullpage}
\usepackage{amsfonts}
\usepackage{graphicx}
\usepackage{mathrsfs}
\usepackage{latexsym}	
\usepackage{oldgerm}
\usepackage{color}
\usepackage{cite}
\usepackage{float}
\usepackage{epstopdf}

\newtheorem{definition}{Definition}[section]

\newtheorem{theorem}{Theorem}[section]
\newtheorem{example}{Example}[section]

\newcommand{\bdf}{\begin{definition}}
\newcommand{\edf}{\end{definition}}
\newcommand{\bthm}{\begin{theorem}}
\newcommand{\ethm}{\end{theorem}}
\newcommand{\bex}{\begin{example}}
\newcommand{\eex}{\end{example}}

\newcommand{\mcal}{\mathcal}

\newcommand{\adag}{a^\dagger}
\newcommand{\bzdag}{b^{(0) \dagger}}
\newcommand{\bdag}{b^\dagger}

\newcommand{\Jint}{\langle \mcal J \rangle_\mrm{int}}
\newcommand{\Jdir}{\langle \mcal J \rangle_\mrm{dir}}

\newcommand{\fr}{\frac}

\newtheorem{prop}{PROPOSITION}[section]
\newcommand{\bprop}{\begin{prop}}
\newcommand{\eprop}{\end{prop}}
\newcommand{\bdis}{\begin{displaymath}}
\newcommand{\edis}{\end{displaymath}}
\newcommand{\beqn}{\begin{equation}}
\newcommand{\eeqn}{\end{equation}}

\newcommand{\eps}{\epsilon}

\newcommand{\mrm}{\mathrm}
\newcommand{\oc}{\omega}
\newcommand{\ok}{\omega_k}

\newcommand{\tret}{\tau_{\mrm{ret}}}

\newcommand{\pt}{\tau}

\newcommand{\bgb}{\begin{gbox}}
\newcommand{\egb}{\end{gbox}}
\newcommand{\bdb}{\begin{darkbox}}
\newcommand{\edb}{\end{darkbox}}

\newcommand{\bsplit}{\begin{split}}
\newcommand{\esplit}{\end{split}}

\usepackage{xcolor}
\usepackage{amsthm}
\usepackage{framed}
\usepackage{authblk}

\colorlet{shadecolor}{gray!15}
\newenvironment{gbox}
  {\begin{shaded}}
  {\end{shaded}}

\title{Quantum Optics of an Oscillator Falling into a Black Hole} 

\date{\today}

\author{ Derek Raine, Paul G. Abel \\
{\small  Department of Physics \& Astronomy,} \\
{\small  University of Leicester, Leicester UK.} \\
{\small  LE1 7RH.} \\
{\small {Email:} pga3@le.ac.uk}}

\begin{document}

\maketitle

\begin{abstract}

We present a quantum optics treatment of the near horizon behaviour of a quantum oscillator freely-falling into a pre-existing  Schwarzschild black hole. We use Painlev\'{e}-Gullstrand coordinates to define a global vacuum state.  In contrast to an accelerated oscillator in the Minkowski vacuum, where there is no radiation beyond an initial transient,  we find that the oscillator radiates positive energy to to infinity and negative energy into the black hole as it attempts to come into equilibrium with the ambient vacuum. We discuss the relationship of the model to Hawking radiation. 

\end{abstract}

\section{Introduction}

Hawking’s original paper \cite{hawking1975particle} showed that when quantum effects are considered, black holes radiate a thermal flux of particles. Despite multiple derivations the physical understanding of Hawking radiation is far from complete with a number of proposed mechanisms include tidal forces on virtual particle-anti-particle pairs analogous to pair creation in an electric field, the splitting of entangled modes as the horizon forms, and quantum tunnelling through the horizon\cite{BROUT1995329} \cite{universalitybh}.
\\
 
Since the Hawking radiation is in a sense universal, independent of details of the collapse phase of matter in the formation of the black hole, it should be possible to understand some of its features by constructing physical models. One such model is the Unruh effect,  introduced in \cite{unruh1976notes} and \cite{davies1975scalar}.  Essentially the idea is to exploit the analogy between a constantly accelerating observer, whose worldline is confined to the Rindler wedge of Minkowski spacetime, and a near- horizon observer in Schwarzschild spacetime at constant radial distance. It is claimed that both detect radiation with a blackbody spectrum at a temperature proportional to the acceleration of the observer as measured at infinity.  Grove \cite{grove1986inertial} was the first to object to this interpretation and suggested instead that the accelerating oscillator emits negative energy with respect to the Minkowski vacuum, which balances out the positive energy emitted by the oscillator as it makes a downward transition, and hence overall there is no net energy flux in the Minkowski vacuum, thereby breaking the analogy. 
\\

This argument was extended further in \cite{raine1991does} and \cite{ford2006there}. These authors also consider a quantum oscillator uniformly accelerating in Minkowski spacetime. To operationalise the meaning of radiation in this context they look at the excitation of a distant inertial detector.  The authors show that the second order fluctuations induced in the field at the detector by the in-falling oscillator balance the first order perturbation exactly and the detector would therefore register no radiation. The result arises because the oscillator and detector are coupled to the same vacuum field. To be clear, in any start-up phase the accelerated oscillator will emit radiation as it comes into equilibrium with the ambient vacuum, but there is no radiation in the steady state.     
\\

It is of obvious interest to extend this model to in-falling quantum systems in a true black hole vacuum. In \cite{scully2018quantum}, the authors examine the transient from a two-level atom falling in the Boulware vacuum. They find a blackbody spectrum arising from the initial response of the atom as a result of the detailed form of the time-dependence along the in-falling trajectory. Because no account is taken of the reaction back on the atom as it decays, the use of first order perturbation theory here is valid on a timescale less than the decay time of an excited state \cite{louisell73quantum}. 
\\

In this paper we investigate a model oscillator falling freely into a black hole as it attempts to come into equilibrium with the ambient vacuum of a scalar field. In 1+1 dimensions we can solve this problem exactly. We use Painlev\'{e}-Gullstrand coordinates  (as in \cite{schutzhold2001hawking}, \cite{PhysRevLett.85.5042} \cite{kanai2012hawking}), with a regular future horizon and future region interior to the horizon, but with an incomplete manifold in the past. The coordinates have the useful feature of a global time which is also the proper time of the in-falling oscillator. This  allows us to define a global vacuum based on the incoming and outgoing modes, with positive and negative frequencies defined everywhere with respect to the proper time.  There is no collapse phase, but the metric is not time-symmetric. We find that in this vacuum the difference between the ingoing and outgoing modes leads to blackbody radiation from the oscillator beyond any contribution from the initial transient.  In contrast in the Boulware vacuum, unsurprisingly because it is static, there is no influence on a distant inertial detector beyond the initial transient. 
\\

The difference between this case and that of the constantly accelerated atom arises from the lack of time-symmetry. Even though the exterior region is static, the outgoing and ingoing wave modes differ. The infalling atom is sensitive to this difference in that it transforms the ingoing modes into a complex mixture of positive and negative frequency outgoing modes. In this respect it acts like reflection in the origin in the original Hawking calculation.  Nevertheless, since the flux at infinity depends on the coupling constant between the atom and the field, this radiation is not the Hawking radiation. Indeed, there is no true Hawking radiation in this set-up. 
\\

The model does however possess some interesting features and possibly tantalising hints. First the radiation here is not the result of the Unruh effect. Indeed we can isolate the "Unruh" terms in the oscillator Hamiltonian -- those corresponding to excitation of the oscillator accompanied by emission of a quantum of the scalar field -- and show that these are a factor (oscillator period)/(decay time) smaller than the dominant (energy conserving) terms. 
\\

We note that the radiation comes from the vicinity of the black hole, but not from within a radial distance from the hole of order $c/\omega$ where $\omega$ is the natural frequency of the oscillator, hence from a distance much greater than the Planck length. Thus, there is no trans-Planckian problem in this model.   
\\

We show that as well as emitting to infinity, the infalling oscillator emits negative energy into the black hole.The Hawking radiation proper arises only when we consider the collapse phase in the formation of a black hole. But this phase is nothing other than the in-fall of a collection of radiating atoms (or oscillators). Thus, the infalling matter emits negative energy into the (forming) black hole. In the Hawking picture the negative energy flux into the hole accompanies the positive energy flux to infinity: these are two parts of the same process. For the infalling atom the ingoing flux perturbs the hole in addition to accompanying the outgoing radiation. The situation is therefore similar to the familiar "burning paper" \cite{preskill1992black}.
\\

The energy going into the hole will perturb the hole and induce it to emit further radiation which will be correlated with the outgoing flux from the infalling matter. If this all happens on an infall timescale, or on the relaxation timescale of the event horizon,  then this is a transient from the collapse  that just happens to have a blackbody form at the Hawking temperature and has little directly to do with Hawking radiation. However, it is intriguing to consider that the information paradox might be resolved if black hole emission is a two-(real)-photon process, linking the emission from the collapsing matter with the evaporating horizon. To decide we need a more detailed model of collapse and evaporation that will allow us to calculate the relaxation time in the presence of the external radiation. 

\section{Gravitational Collapse in Painlev\'{e}-Gullstrand}

We consider a quantum oscillator falling into a Schwarzschild black hole in $1+1$ dimensions.    The  Painlev\'{e}-Gullstrand coordinate system provides a convenient framework since the coordinates are regular in the exterior region and across the future horizon; the time coordinate is also conveniently the proper time of the infalling oscillator. We adopt natural units: $\hbar=G=c=1$.
\\

\begin{figure}
%\centerline{\fbox{\includegraphics[scale=0.75]{figs/Penrose_SADS.eps}}}
\centering
\includegraphics[width=0.5\textwidth]{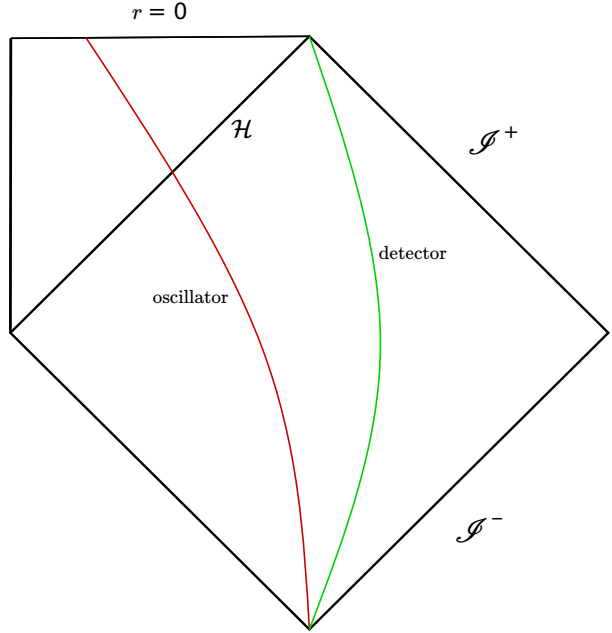}
\caption{The Penrose-Carter diagram showing the oscillator on a free-fall trajectory in Painlev\'{e}-Gullstrand coordinates.  
The detector remains outside the black hole.}
\label{Penrose-Carter}
\end{figure}

In terms of Schwarzschild coordinates $(t_s, r)$ the Panlev\'{e}-Gullstrand time $\tau$ is given by $\tau =t_s -h(r)$, where the function $h(r)$ is obtained from
\beqn
\label{PG-a}
\frac{dh}{dr} = \mp \left(1-\frac{2M}{|r|} \right)^{-1}\sqrt{\frac{2M}{|r|}}.
\eeqn
In $1+1$ dimensions the spatial coordinate $r$ ranges over $-\infty < r < +\infty$ but we shall be concerned with only the region $r>0$, so we assume that $r$ is positive throughout.  Adopting these coordinates, the 1+1 metric for a black hole of mass $M$  becomes
\beqn
-ds^2=-d\pt^2+\left(dr + \sqrt{\frac{2M}{r}} d\pt\right)^2.
\eeqn
The equation of motion of a particle falling freely from rest at infinity is {\cite{kanai2012hawking}}
\beqn
r(\pt)=\left(\fr{9M}{2}(-\pt)^2\right)^{1/3},
\eeqn
where, writing $\tau_s =4M/3$, in the exterior region $-\infty < \tau \leq - \tau_s $, and $\tau_s < \tau <0$ in the black hole interior.
As in \cite{raine1991does} in order to operationalise the existence of radiation from the infalling oscillator, we place a detector on a world line $r = $constant, at some large distance from the event horizon.  The Penrose-Carter diagram showing this scenario is given in figure \ref{Penrose-Carter}.
The relationship between Schwarzschild time $t_s$ and Painlev\'{e}-Gullstrand coordinates $(\pt,r)$ is:
\beqn
dt_s=d\pt-\fr{\sqrt{2M/r}}{1-2M/r}dr.
\eeqn
In terms of the usual tortoise coordinate
\beqn
r_*=r+2M\ln\left|\fr{r}{2M}-1 \right|,
\eeqn
for $r > 0$ we define the \emph{out-going} null coordinates  
\beqn
u=t_s-r=\pt-\xi(r),
\eeqn
with
\beqn
\xi(r)=r+2\sqrt{2Mr}+4M\ln\left(\sqrt{\fr{r}{2M}}-1 \right),
\eeqn
and similarly, the \emph{in-going} null coordinate for $r > 0$,
\beqn
v=t_s+r_s=\pt+\eta(r),
\eeqn
with
\beqn
\eta(r)=r-2\sqrt{2Mr}+4M\ln\left(\sqrt{\fr{r}{2M}} + 1\right).
\eeqn

We couple the oscillator to a massless scalar field $\Phi$ satisfying the Klein-Gordon equation,
\beqn
\Box\Phi=0.
\eeqn
In P-G coordinates in 1+1 dimensions this is
\beqn
(1-f)^{-1}\left[\frac{\partial}{\partial t} - (1-f) \frac{\partial}{\partial r} \right]\left[ (1-f) \frac{\partial}{\partial t} + (1- f^2) \frac{\partial}{\partial r}\right] = 0
\eeqn
where $f= \sqrt{2M/r}$.
In the right hand wedge the outgoing field can be expanded  in terms of out-going modes $e^{\pm i k u}$ 
\beqn
\phi_{\mrm{out}}=\sum_{k\geq 0}e_k \left\{b^{(0)}_k e^{-ik(t-\xi(r))}+\bzdag_{k}e^{ik(t-\xi(r)}\right\}
\eeqn
with the usual Klein-Gordon normalisation in a box of length $L$, $e_k = \sqrt{\frac{2\pi}{kL}}.$ 
%%%
\noindent Similarly we can express the field in terms of in-going modes $ e^{ \pm i k v}$ in the wedge $r>0$ as 
\beqn
\label{Phi in} 
\phi_{\mrm{in}}=\sum_{k\geq 0}e_k\left\{b^{(0)}_{-k}e^{-ik(t+\eta(r))}+\bzdag_{-k}e^{ik(t+\eta(r))}\right\}.
%\phi_{\mrm{out}}=\int_{0}^{\infty}\frac{dk}{\sqrt{4\pi k}}\left\{b_{-k}e^{-ik(t+\eta(r))}+\bdag_{-%k}e^{ik(t+\eta(r))}\right\}.
\eeqn

\noindent In general, the full solution to the Klein-Gordon equation will be a linear combination of the ingoing and 
outgoing modes:
\beqn
\Phi=\phi_{\mrm{in}}+\phi_{\mrm{out}}.
\eeqn
We define the vacuum state $|0\rangle$ by
\beqn
b^{(0)}_k|0\rangle = 0 \quad \text{and} \quad b^{(0)}_{-k}|0\rangle=0.
\eeqn

We shall need the Fourier transforms of these modes as evaluated along the worldline of the infalling oscillator.  We define
\beqn
e^{ik(\pt-\xi(\pt))}=\left\{\begin{array}{l l}\displaystyle{\int_{-\infty}^\infty} \alpha_k(k') e^{-ik'\pt} \ dk' & \quad \tau < - \tau_s \\
			0  & \quad \tau > - \tau_s\\
\end{array} \right. 
\label{alpha}
\eeqn
and
\beqn
e^{ik(\pt+\eta(\pt))}=\left\{\begin{array}{l l} \displaystyle{\int_{-\infty}^\infty} \beta_k(k')e^{-ik'\pt}\ dk' & \quad \tau < - \tau_s \\
				0  & \quad \tau > - \tau_s.\\
\end{array} \right. 
\label{beta}
\eeqn

\noindent Approximate expressions for the Fourier components $\alpha_k(k')$ and $\beta_k(k')$ are obtained in appendix 1. We find 
\begin{align}
\alpha_k(k')&= \fr{2M}{\pi}ke^{- 2 \pi Mk}e^{-iM(22k+4k')/3}(4M)^{4iMk}(3k+k')^{4iMk-1}\Gamma(-4iMk)\label{Alpha1}\\
\beta_k(k') &=2 \sqrt{\frac{2 M}{\pi|k|}}e^{i\pi /4}\exp\left\{ iMk\left[-\frac{10}{3}+4 \ln 2\right] +ik'\tau_s-\frac{2iM}{k}(k+2k')^2 \right\} . \label{beta1}
\end{align}
If $3k + k'$ is real, then the branch in the imaginary power is defined by $3k+k' = e^{- i \pi} |3k+k'|$ for $3k+k' < 0.$ 

%%%%%%%%%%%%%%%%%%%%%%%
\section{The Quantum Langevin Equation}

We now derive the equation of motion for the oscillator.  We take a quantum harmonic oscillator of mass $m$ and natural
frequency $\oc$  confined to a free-fall worldline in $r > 0$  with  proper time $\pt$. We couple this to a massless scalar field $\Phi$ with a scalar-electrodynamic form for the interaction. Our Hamiltonian is therefore
\beqn
\begin{split}
\mathscr H =\oc \adag a+&\sum_{k>0} \ok \bdag_kb_k  + \sum_{k>0} \ok \bdag_{-k}b_{-k}\\
&ig\sqrt{\frac{\omega}{2m}} \sum_{k>0} e_k(\adag-a)(b_ke^{ik\xi}+\bdag_ke^{-ik\xi}+b_{-k}e^{-i\ok \eta}+\bdag_{-k}e^{ik\eta})
\end{split}
\label{hamiltonian}
\eeqn
where $\adag$ is the creation operator for the quantum oscillator, $\ok = |k|$, and $g$ is a coupling constant.
\\

We do not make the rotating wave approximation at this point \cite{louisell73quantum} to remove the products that pair creation operators of the field and oscillator (and similarly pairings of annihilation operators) both because keeping them here makes the calculation slightly easier and for comparison with the Unruh radiation in Scully et al. \cite{scully2018quantum} later. We shall impose the rotating wave approximation appropriately below. 
\\

We now use Heisenberg's equation of motion to determine the evolution of the oscillator and the scalar field. 
From
\beqn
\fr{da}{d\pt}=-i[a,\mathscr H], 
\eeqn
and putting $\lambda = g \sqrt{\omega / 2m} $ for brevity, we obtain the equation of motion for the oscillator
\beqn
\fr{da}{d\pt}=-i\oc a +\lambda \sum_k e_k(b_ke^{ik\xi}+\bdag_ke^{-ik\xi}+b_{-k}e^{-i\ok \eta}+\bdag_{-k}e^{ik\eta}).
\label{osc1}
\eeqn
Similarly, for the scalar field:
\beqn
\fr{db_j}{d\pt}=-i[b_j,\mathscr H],
\eeqn
from which we obtain
\beqn
\fr{db_k}{d\pt}=-i\ok b_k+\lambda e_k(\adag-a)e^{-ik\xi} \ \ \mathrm{and} \ \ 
\fr{db_{-k}}{d\pt}=-i\ok b_{-k}+\lambda e_k(\adag-a)e^{ik\eta}.
\label{field1}
\eeqn

\noindent We can solve (\ref{field1}) to obtain expressions for the scalar field  operators:
\beqn
b_k(\pt)=e^{-ik\pt}b_k^{(0)}+\lambda e^{-ik\pt}e_k\int_{-\infty}^\pt (\adag-a)e^{ik(\pt'-\xi(\pt'))}\ d\pt'
\label{bk}
\eeqn
and
\beqn
b_{-k}(\pt)=e^{-ik\pt}b_{-k}^{(0)}+\lambda e^{-ik\pt}e_k\int_{-\infty}^\pt (\adag-a)e^{ik(\pt'+\eta(\pt'))}\ d\pt'.
\label{b-k}
\eeqn
We assume that the interaction is switched on at some distant time $\tau > -\infty$  in the past when $\lambda \rightarrow 0$ and $b_k = b_k^{(0)}.$   
\\

We now go on to use (\ref{bk}) and (\ref{b-k}) to derive an expression for the position operator for our oscillator. Direct
substitution into (\ref{osc1}) gives:
\beqn
\begin{split}
\fr{da}{d\pt}=&-i\oc a+\mathcal G_a +\lambda^2\sum_k e_k^2 \bigg\{  \\
&e^{-ik(\pt-\xi)}\int_{-\infty}^\pt(\adag-a)e^{ik(\pt'-\xi')}\ d\pt'
-e^{ik(\pt-\xi)}\int_{-\infty}^\pt(\adag-a)e^{-ik(\pt'-\xi')}\ d\pt'\\
&+e^{-ik(\pt+\eta)}\int_{-\infty}^\pt(\adag-a)e^{ik(\pt'+\eta')}\ d\pt'
-e^{ik(\pt+\eta)}\int_{-\infty}^\pt(\adag-a)e^{-ik(\pt'+\eta')}\ d\pt'\bigg\}
\end{split}
\label{osc2}
\eeqn
with the function
\beqn
\mathcal G_a(\pt)=\lambda \sum_k e_k\{ b_{k}^{(0)}e^{-ik(\pt-\xi)}+\bzdag_{k}e^{ik(\pt-\xi)} 
+b_{-k}^{(0)}e^{-ik(\pt+\eta)}+\bzdag_{-k}e^{ik(\pt+\eta)}\}.
\label{Ga}
\eeqn
We now remove high frequency behaviour by setting
\beqn
a(\pt)=e^{-i\oc \pt}A(\pt).
\label{A}
\eeqn
Using this and the Fourier transforms of (\ref{alpha}) and (\ref{beta}) in (\ref{osc2}) gives:
\beqn
\begin{split}
\fr{dA}{d\pt}=&\mathcal G_A+\lambda^2\sum_k e_k^2 e^{i\oc\pt} \bigg\{ \\
&\int_{-\infty}^\infty \alpha^*(k'')e^{ik''\pt}\ dk''\int_{-\infty}^\infty\alpha_k(k')\ dk' \int_{-\infty}^\pt\left(
e^{i(\oc-k')\pt'}A^\dagger-e^{-i(\oc+k')\pt' }A\right)\ d\pt' \\
&-\int_{-\infty}^\infty \alpha(k')e^{-ik'\pt}\ dk'\int_{-\infty}^\infty\alpha_k^*(k'')\ dk'' \int_{-\infty}^\pt\left(
e^{i(\oc+k'')\pt'}A^\dagger-e^{-i(\oc-k'')\pt' }A\right)\ d\pt' \\
&+\int_{-\infty}^\infty \beta^*(k'')e^{ik''\pt}\ dk''\int_{-\infty}^\infty\beta_k(k')\ dk' \int_{-\infty}^\pt\left(
e^{i(\oc-k')\pt'}A^\dagger-e^{-i(\oc+k')\pt' }A\right)\ d\pt' \\
&-\int_{-\infty}^\infty \beta(k')e^{-ik'\pt}\ dk'\int_{-\infty}^\infty\beta_k^*(k'')\ dk'' \int_{-\infty}^\pt\left(
e^{i(\oc+k'')\pt'}A^\dagger-e^{-i(\oc-k'')\pt' }A\right)\ d\pt' \bigg\}.
\end{split}
\label{osc-3a}
\eeqn
The rotating wave approximation now amounts to neglecting the $A^\dagger$ terms since these behave like $e^{2\oc \pt}$. We could subsequently include these terms in an iterative solution, which would amount to taking into account the  higher energy levels of the oscillator in the line profile, but this is not crucial for our discussion. (In effect we are treating the oscillator as a two-level atom.) 
\\

Next we perform the $\pt'$ integrations using integration by parts:
\beqn
\int_{-\infty}^\pt e^{-i(\oc+k')\pt'}A(\pt')d\pt' =
\left[\fr{i A(\pt')e^{-i(\oc+k')\pt'}}{\oc+k'} \right]_{-\infty}^\pt-i\int_{-\infty}^\pt
\fr{dA}{d\pt'}\fr{e^{-i(\oc+k')\pt'}}{\oc+k'}\ d\pt'
\label{IBP}
\eeqn  
We can neglect
the final (integral) term in (\ref{IBP}) because, from (\ref{osc-3a}), it contributes a correction of order $\lambda^2$ to $\mathcal G_A $ and of order $\lambda^4$ to $dA/d \pt$. This method is equivalent to solving equation (\ref{osc-3a}) via a Laplace transform as  in Louisell \cite{louisell73quantum}. 
(We demonstrate this equivalence in appendix 5.)  The contribution from $\tau' \rightarrow -\infty$ vanishes  under our assumption that the interaction is switched off at early times. We are left with
\beqn
\begin{split}
\fr{dA}{d\pt} &= \mathcal G_A+i\lambda^2 A(\pt)\sum_k e_k^2 \bigg\{  \\
&-\int_{-\infty}^\infty \alpha^*(k'')e^{ik''\pt}dk'' \int_{-\infty}^\infty \fr{\alpha_k(k')e^{-ik'\pt}dk'}{\oc+k'}
+\int_{-\infty}^\infty \alpha(k')e^{-ik'\pt}dk' \int_{-\infty}^\infty \fr{\alpha_k^*(k'')e^{ik''\pt}dk''}{\oc-k''}\\
&-\int_{-\infty}^\infty \beta^*(k'')e^{ik''\pt}dk'' \int_{-\infty}^\infty \fr{\beta_k(k')e^{-ik'\pt}dk'}{\oc+k'}
+\int_{-\infty}^\infty \beta(k')e^{-ik'\pt}dk' \int_{-\infty}^\infty \fr{\beta_k^*(k'')e^{ik''\pt}dk''}{\oc-k''} \bigg\}.
\end{split}
\label{osc3}
\eeqn

This expression can  be simplified by interchanging $k'$ and $k''$ in the second integrals on each of the last two lines of equation (\ref{osc3}) and using $\alpha_k(k') = \alpha_{-k}^{*}(-k')$ with a similar relation for $\beta_k(k')$ to combine the integrals into a sum over $k$, $-\infty < k < \infty$. Finally, writing the sum over $k$ as an integral (i.e. letting $L \rightarrow \infty$) we get  %% Louiselle p250, para 4.5

\beqn
\begin{split}
\fr{dA}{d\pt} &= \mathcal G_A \\ 
& -i\lambda^2 A(\pt)\int_{-\infty}^{\infty} \frac{dk}{k}\int_{-\infty}^{\infty}dk' \int_{-\infty}^{\infty}dk'' \left[ \alpha_{k}^{*}(k'') \alpha_k(k') + \beta^{*}_k(k'') \beta_k(k') \right] \frac{e^{i(k''-k')\pt}dk'}{\oc+k'}
\end{split}
\label{q-lang-aa-bb}
\eeqn
Defining ${\mathcal G_A} = e^{i\oc \tau}\mathcal G_a$, we write the quantum Langevin equation for the oscillator as
\beqn
\fr{dA}{d\pt}=-\left(\fr{\gamma}{2}+ i\Delta \omega\right)A(\pt)+\mathcal G_A,
\label{q-langevin}
\eeqn
with the friction constant $\gamma$ and the Lamb shift $\Delta \omega$ implicitly given by (\ref{q-lang-aa-bb}). 
\\

We show in appendix 3 that

\beqn
\gamma = \frac{ \pi g^2}{m}. 
\label{gamma1}
\eeqn 

\noindent The evaluation of the Lamb shift is more subtle and will be pursued elsewhere. Here we simply  incorporate it into the definition of $\oc$.

Solving (\ref{q-langevin}) we get
\beqn
A(\pt)=e^{-\gamma\pt/2}\int_{-\infty}^\pt e^{\gamma\pt'/2}\mathcal G_A(\pt') \ d\pt' 
\eeqn
where we have ignored the initial value of $A$ since this gives rise to a transient signal far from the black hole and is consequently of no interest here. 
%where from (\ref{Ga}) we have that
%\beqn
%G_A(\pt)=\lambda e^{i\oc \pt}\sum_k e_k\left\{ 
%b_{k}^{(0)}e^{-ik(\pt-\xi)}+\bzdag_{k}e^{ik(\pt-\xi)}+b_{-k}^{(0)}e^{-ik(\pt+\eta)}+\bzdag_{-k}e^{ik(\pt+\eta)} %\right\}.
%\eeqn

Thus,
\beqn
\begin{split}
A(\pt)& =\\
&\lambda e^{-\gamma \pt/2} \sum_k e_k\left\{ 
b_{k}^{(0)}\int_{-\infty}^\pt e^{i(\oc-i\gamma/2)\pt'}e^{-ik(\pt'-\xi(\pt'))}d\pt'+\bzdag_{k}\int_{-\infty}^{\pt}
e^{i(\oc-i\gamma/2)\pt'}e^{ik(\pt'-\xi(\pt'))}\ d\pt'\right.\\
&\left. +b_{-k}^{(0)}\int_{-\infty}^\pt e^{i(\oc-i\gamma/2)\pt'}e^{-ik(\pt'+\eta(\pt'))}d\pt'+\bzdag_{-k}\int_{-\infty}^{\pt}
e^{i(\oc-i\gamma/2)\pt'}e^{ik(\pt'+\eta(\pt'))}\ d\pt'\right\}.
\end{split}
\eeqn
Using the Fourier transforms (\ref{alpha}) and (\ref{beta}) gives:
\beqn
\begin{split}
A(\pt)&=\lambda e^{-\gamma \pt/2} \sum_k e_k \bigg\{ \\
&b_{k}^{(0)}\int_{-\infty}^\infty\int_{-\infty}^\pt e^{i(\oc+k'-i\gamma/2)\pt'}\alpha^*(k')d\pt'dk'+
\bzdag_k\int_{-\infty}^\infty\int_{-\infty}^\pt e^{i(\oc-k'-i\gamma/2)\pt'}\alpha(k')d\pt'dk'\\
&+b_{-k}^{(0)}\int_{-\infty}^\infty\int_{-\infty}^\pt e^{i(\oc+k'-i\gamma/2)\pt'}\beta^*(k')d\pt'dk'+
\left.\bzdag_{-k}\int_{-\infty}^\infty\int_{-\infty}^\pt e^{i(\oc-k'-i\gamma/2)\pt'}\beta(k')d\pt'dk' \right\}.
\end{split}
\eeqn
We define the oscillator susceptibility
\beqn
\chi(k')=\frac{1}{\oc+k'-i\gamma/2}.
\label{chi}
\eeqn
 Performing the $\pt'$ integrations we obtain 
\beqn
\begin{split}
A(\pt)&=\\
& -i\lambda e^{i\oc \pt}\sum_k e_k \left\{
b_k^{(0)}\int_{-\infty}^\infty e^{ik'\pt} \alpha^*(k')\chi(k')dk'+\bzdag_k\int_{-\infty}^\infty e^{-ik'\pt}\alpha_k(k')\chi(-k')dk'\right.\\
&\left.+b_{-k}^{(0)}\int_{-\infty}^\infty e^{ik'\pt}\beta_{k}^*(k')\chi(k')dk'+\bzdag_{-k}\int_{-\infty}^\infty e^{-ik'\pt}\beta_k(k')\chi(-k')dk'
\right\}.
\end{split}
\eeqn
In principle, this would allow us to determine how the oscillator comes into equilibrium with its local environment at any distance from the black hole. However, we know the $k'$-dependence of $\alpha_{k}(k')$ and $\beta_{k}(k')$ only on the assumption that $\pt < \sim \tau_s$.  
\\

We now have everything we require to determine the position operator of the oscillator
\beqn
q=\fr{1}{\sqrt{2\oc m}}(\adag+a)
\eeqn
near the black hole.
Using  $A(\pt)=e^{i\oc\pt}a(\pt)$, and writing
\beqn
\Delta\chi(k') = \chi^*(k') - \chi(-k'),
\label{Delta chi}
\eeqn 
the position operator of the harmonic oscillator, ignoring transients, becomes
\beqn
\begin{split}
q(\pt)=\fr{i\lambda}{\sqrt{2m\oc}}&\sum_k e_k \bigg\{ \\
 & b_k^{(0)}\int_{-\infty}^\infty e^{ik'\pt}\alpha^*(k')\Delta \chi(-k') dk'
+\bzdag_k\int_{-\infty}^\infty e^{-ik'\pt}\alpha_k(k')\Delta \chi (k')dk'\\
+ & \left. b_{-k}^{(0)}\int_{-\infty}^\infty e^{ik'\pt}\beta^*(k')\Delta \chi(-k')  dk'
+\bzdag_{-k}\int_{-\infty}^\infty e^{-ik'\pt}\beta_k(k')\Delta \chi(k')dk'\right\}.
\end{split}
\label{q}
\eeqn

%%%%%%%%%%%%%%    FIELD EQUATIONS       %%%%%%%%%

\section{The Solution to the Field Equation}

We now wish to determine the solution to the scalar field equation in the presence of the oscillator.  In section 2 we determined that the scalar field
$\Phi$ can be decomposed into a linear sum of in-going and outgoing modes:
\beqn
\Phi=\sum_k e_k \left\{
b_k e^{ik\xi}+\bdag_ke^{-ik\xi}+b_{-k}e^{-ik\eta}+\bdag_{-k}e^{ik\eta}
\right\}.
\label{Phi1}
\eeqn 
In section 3 we determined expression for $b_k$ and $b_{-k}$; these are given in (\ref{bk}) and (\ref{b-k}). Thus substituting
into (\ref{Phi1}) we find that the field can be written as
\beqn
\Phi=\Phi_h+\Phi_p,
\eeqn
where the homogeneous part
\beqn
\Phi_h=\sum_k e_k\left\{
b_k^{(0)}e^{-ik(\pt-\xi)}+b_{k}^{\dagger (0)}e^{ik(\pt-\xi)}+b_{-k}^{(0)}e^{-ik(\pt+\eta)}+b_{-k}^{\dagger (0)}e^{-ik(\pt+\eta)} \right\},
\eeqn
and the particular integral is
\beqn
\begin{split}
\Phi_p=&\lambda \sum_k e_k^2 \Big\{  \\
&e^{-ik(\pt-\xi(\pt))}\int_{-\infty}^\pt(\adag-a)e^{ik(\pt'-\xi(\pt'))} \ d\pt'
-e^{ik(\pt-\xi(\pt))}\int_{-\infty}^\pt(\adag-a)e^{-ik(\pt'-\xi(\pt'))} \ d\pt' \\
&e^{-ik(\pt+\eta(\pt))}\int_{-\infty}^\pt(\adag-a)e^{ik(\pt'+\eta(\pt'))} \ d\pt'
-e^{ik(\pt+\eta(\pt))}\int_{-\infty}^\pt(\adag-a)e^{-ik(\pt'+\eta(\pt'))} \ d\pt'\Big\}.
\end{split}
\eeqn
Using the relation between the annihilation and creation operators and the momentum, $p$,
\beqn
 p=m\frac{dq}{d\tau} = i\sqrt{\frac{m \oc}{2}}  (a^{\dagger} - a),
\eeqn
we get
\beqn
\begin{split}
\Phi_p=-ig&\sum_ke_k^2\left\{ \int_{-\infty}^\pt \fr{dq}{d\pt'}\left( e^{ik(\pt'-\pt)-ik(\xi(\pt')-\xi(\pt))}-
e^{-ik(\pt'-\pt)+ik(\xi(\pt')-\xi(\pt))}\right)\ d\pt' \right. \\
&+\int_{-\infty}^\pt\fr{dq}{d\pt'}
\left.\left(e^{ik(\pt'-\pt)+ik(\eta(\pt')-\eta(\pt))}-e^{-ik(\pt'-\pt)-ik(\eta(\pt')-\eta(\pt))} \right)\ d\pt'\right\}.
\end{split}
\eeqn
Combining the exponentials and converting $\sum_k\rightarrow \int dk$ gives
\beqn
\begin{split}
\Phi_p=2g&\left[ 
\int_0^\infty \int_{-\infty}^\pt \fr 1 k \fr{dq}{d\pt'} \sin[k(\pt'-\pt)-k(\xi(\pt')-\xi(\pt)]\ dt\ dk \right. \\
&\left.+\int_{-\infty}^0 \int_{-\infty}^\pt \fr 1 k \fr{dq}{d\pt'} \sin[k(\pt'-\pt)-k(\eta(\pt')+\eta(\pt)]\ dt\ dk
\right].
\end{split}
\eeqn
We now set
\beqn
I_1=\int_0^\infty \int_{-\infty}^\pt\fr 1 k \fr{dq}{d\pt'}\sin[k(\pt'-\pt)-k(\xi(\pt')-\xi(\pt))]\ d\pt'dk, 
\eeqn
and
\beqn
I_2=\int_{-\infty}^0 \int_{-\infty}^\pt\fr 1 k \fr{dq}{d\pt'}\sin[k(\pt'-\pt)+k(\eta(\pt')+\eta(\pt))]\ \pt'dk.
\eeqn
Evaluating the $k$ integral in $I_1$ first, we see that this is just the retarded Green's function in two dimensions
\cite{raine1991does}, and so:
\beqn
\mathscr G_{\mathrm{ret}}(\pt, \xi, \pt', \xi')=
\left\{ \begin{array}{cc} \pi & |\xi-\xi'|<\pt-\pt', \ \pt>\pt' \\ 0 & \mathrm{otherwise} \end{array} \right.
\eeqn
which means that
\beqn
I_1=q(\pt_{\mathrm{ret}})
\eeqn
where $\pt_{\rm ret}$ is given by
\beqn
\label{tau-ret}
\pt_{\rm ret}-\xi(\pt_{\rm ret})= \pt'-\xi'.
\eeqn
Evaluating $I_2$ we identify the $k$ integral in $I_2$ as being the advanced Green's function, and if we let
$k\rightarrow -k$ 
\beqn
\mathscr G_{\mathrm{adv}}=\lim_{\eps\rightarrow 0} \left\{ \fr{1}{2\pi}
\int_0^\infty \int_{-\infty}^\infty \fr{e^{ik(\eta-\eta')-ik^0(\pt-\pt')}}{(k^0-i\eps)^2-k^2} \ dk^0 dk.\right\}
\eeqn
Examining the $k^0$ integral we see that the simple poles are located in the upper-half of the complex plane. However we still require $\pt>\pt'$ so we would form a semi-circular in the lower half of the complex plane which
does not therefore enclose the poles.  So this integral gives no contribution. Thus we have obtained the
solution to the scalar field equation
\beqn
\Phi(\pt,r)=\Phi_h+2\pi g q(\tret).
\label{Phi}
\eeqn

%%%%%%%          ENERGY FLUX          %%%%%%%%%%%%%%%%%%

\section{The Energy Flux at the Detector}

We now look at the response of a detector at a large distance from the black hole. This operationalises the meaning of radiation from the infalling oscillator. For the accelerated detector in the Rindler wedge in \cite{raine1991does} we calculated the noise power on the world line of a distant inertial detector, which is directly related to the probability of excitation of the detector. Here we shall look at the closely related, but more familiar,  energy flux at the detector. 
\\

We shall demonstrate first that the energy flux at the detector has a blackbody form modulated by the impedance of the infalling body.  This enables us to explain the difference between the Rindler case and a black hole. We also show that the oscillator emits a negative energy flux into the hole. We then use the explicit form for the impedance function of a harmonic oscillator to derive an explicit form for the energy flux from the infalling oscillator. 
\\

Let the P-G  coordinates of the oscillator be $(\pt, \xi(\pt))$ and let the coordinates of the detector be $(t ', \xi'(t'))$. For $\xi' > \xi$
\beqn
\pt_{\rm ret} - \xi(\pt_{\rm ret})=t'- \xi',
\eeqn
and for $\xi' < \xi$ 
\beqn
\pt_{\rm ret} + \eta(\pt_{\rm ret}) = t' + \eta'.
\eeqn
 In P-G coordinates an orthonormal dyad in the rest frame of the detector is
\beqn
{\bf e}^{0}=(1, 0), \quad {\bf e}^{1} = (x', 1).
\eeqn
The energy-momentum flux at the detector is 
\beqn
F =T^{\hat{1}}_{\hat{0}} =  - T_{\hat{0}\hat{1}} =- \langle {\bf e}^{\mu}_{0}{\bf e}^{\nu}_{1}T_{\mu \nu} \rangle   = - x'T_{r'r'}+T_{t'r'},
\eeqn
where the components of $T_{\mu' \nu'}$ are obtained as usual from
\beqn
T_{\mu' \nu'} = \frac {\partial \Phi}{\partial x'^{\mu}}\frac{\partial \Phi^{\dagger}}{\partial x'^{\nu}}-\frac{1}{2}g_{\mu' \nu'} \left(\frac{\partial \Phi}{\partial x'^{\lambda}} \right)^2.
\eeqn
Using (\ref{Phi}), the expectation value of the energy momentum flux is obtained from
\beqn
\label{noisy}
\begin{split}
 \left\langle \frac{\partial \Phi^{\dagger}}{\partial x'^{\mu}}  \frac{\partial \Phi}{\partial x'^{\nu}}\right\rangle = & 
\left\langle \frac{\partial \Phi_{h}^{\dagger}}{\partial x'^{\mu}}  \frac{\partial \Phi_h}{\partial x'^{\nu}}\right\rangle  \\ 
&+ 2 \pi g \left\langle \frac{\partial \Phi_{h}^{\dagger}}{\partial x'^{\mu}} \dot{q}(\tau_{ret}) \frac{\partial \tau_{ret}}{\partial x'^{\nu}}+     \dot{q^{\dagger}}(\tau_{ret}) \frac{\partial \tau_{ret}}{\partial x'^{\mu}} \frac{\partial \Phi_h}{\partial x'^{\nu}}\right\rangle  \\ 
&  + 4 \pi g^2 \left\langle \dot{q^{\dagger}}(\tau_{ret}) \frac{\partial \tau_{ret}}{\partial x'^{\mu}}  \dot{q}(\tau_{ret}) \frac{\partial \tau_{ret}}{\partial x'^{\nu}}\right\rangle.
\end{split}
\eeqn
The first term on the right of equation (\ref{noisy}) involves only the unperturbed field and represents the flux present in the absence of the oscillator. The Hawking radiation is obtained by the standard calculation (e.g. \cite {hawking1975particle}) that relates the incoming modes on $\mathscr{I}^{-}$ at $r \rightarrow -\infty$ that do not fall into the horizon (defining the in-vacuum) to the outgoing modes on $\mathscr{I}^{+}$ at $r\rightarrow +\infty$, defining the out-vacuum. In the Painlev\'{e}-Gullstrand  manifold here (figure \ref{Penrose-Carter})   there is only one vacuum state and no mixing of modes in the absence of the infalling oscillator. This term is therefore just the zero point flux and will have no influence on the detector. We can confirm this by an explicit calculation. 
\\

The normally ordered expression for the unperturbed $\langle : T_{\mu' \nu'}: \rangle$  based on the Painlev\'{e}-Gullstrand modes gives zero contribution. The covariant form can be calculated from the conformal factor,  $e^{2\rho}=(1-f^2)$, in the usual way for a 1+1 dimensional metric \cite{bh-evap2005} \cite{davies1976}
\beqn
\langle T_{uu}\rangle  \propto \frac{\partial^2 \rho}{\partial u ^2} -\left(\frac{\partial \rho}{\partial u}\right)^2
\eeqn
with similar expressions for $\langle T_{vv} \rangle$. We obtain (up to a numerical factor) 
\beqn
\langle T_{uu}\rangle= \langle T_{vv}\rangle =\frac{M}{2r^3}- \frac{3M^2}{4r^4}
\eeqn
and
\beqn
\langle T_{uv}\rangle=\frac{\partial^2 \rho}{\partial u \partial v} = - \frac{M}{2r^3}\left( 1 - \frac{2M}{r} \right).
\eeqn
which vanish at infinity (exactly as in the Boulware vacuum). Thus there is no Hawking flux. (The flux on the horizon is formally non-zero, but this is non-physical since the coordinates do not satisfy the regularity conditions there \cite{christensen1977}.)  
Of course, the situation would be different if we were to take into account the collapse phase in the formation of the black hole, when this term would yield the usual Hawking effect.    
\\

The term on the second line of (\ref{noisy})  represents the interference between the outgoing emission from the oscillator and the  vacuum excitations of the detector. This is the flux we would get from first order perturbation theory treating the scalar field as an external  potential. The final term in (\ref{noisy}) represents the direct contribution to the flux arising from the oscillator. 
\\

We can tidy up equation (\ref{noisy}) using the fact that the time dependence in $\Phi$ comes through $\tau_{\rm ret}$, since $\Phi_{h}(t'-\xi') = \Phi_{h}(\tau_{\rm ret} - \xi(\tau_{\rm ret}))$ and $q=q(\tau_{\rm ret})$. 
Thus the derivatives in $T_{\mu\nu}$ contribute factors of:
\begin{eqnarray*}
 \frac{\partial \Phi}{\partial r'}  \frac{\partial \Phi^{\dagger}}{\partial r'}&\rightarrow &  \langle \dot{\Phi} \dot{\Phi}^{\dagger}  \rangle \left(\frac{\partial \tau}{\partial \xi'}\right)^{2} \left( \frac{\partial \xi'}{\partial r'}\right)^2\\
 \frac{\partial \Phi}{\partial r'}  \frac{\partial \Phi^{\dagger}}{\partial t'}& \rightarrow &  \langle \dot{\Phi} \dot{\Phi}^{\dagger}\rangle  \left(\frac{\partial \tau}{\partial \xi'}\right) \left( \frac{\partial \xi'}{\partial r'}\right)\left(\frac{\partial \tau}{\partial t'} \right)\\
\frac{\partial \Phi}{\partial t'}  \frac{\partial \Phi^{\dagger}}{\partial t'}& \rightarrow  &   \langle \dot{\Phi} \dot{\Phi}^{\dagger}\rangle \left(\frac{\partial \tau}{\partial t'}\right)^{2}
\end{eqnarray*}
where $\dot{\Phi} = d\Phi/d\tau_{\rm ret}.$
Now, $q(\tau_{\rm ret})$ contributes a factor $\exp(\mp ik' \tau_{\rm ret})$ to $\Phi$ and $\Phi_h = \phi_{\rm out}$ contributes a factor $\exp[\pm ik'(t'-\xi')] = \exp[\pm ik'(\tau_{\rm ret} - \xi (\tau_{\rm ret}))]$. Thus $\langle \dot{\Phi} \dot{\Phi}^{\dagger} \rangle = k'k'' \langle \Phi \Phi^{\dagger} \rangle$.
Writing $f = \sqrt {2M/r} $, $f'= \sqrt{2M/r'}$  as above (for $r>0$, $r'>0$), from the definitions (\ref{tau-ret}) we find
\beqn
\frac{\partial \tau}{\partial \xi'} =- \frac{\partial \tau}{\partial t'}= -(1-f) \quad \text{and} \quad \frac{\partial \xi'}{\partial r'} = \frac{1}{1-f'}. \quad 
\eeqn
Putting this together we find that
\beqn
F = \frac{(1-f)^2}{(1-f')^2} (\Jdir + \Jint)
\label{Fdir}
\eeqn
where 
\beqn
\label{Jdir1}
\begin{split}
&\Jdir =  4 \pi g^2 \langle \dot{q}^\dagger \dot{q} \rangle \\
=&\gamma^2 \sum_k e_k^2 
\int_{-\infty}^\infty\int_{-\infty}^\infty
k' k'' e^{i(k''-k')\tret}\Delta \chi(k') \Delta \chi^*(k'')       %(\chi^*(k')-\chi(-k'))(\chi(k'')-\chi^*(-k''))
[\alpha_k(k')\alpha_k^*(k'')+\beta_k(k')\beta^*(k'')]\ dk' dk''.
\end{split}
\eeqn
with $\gamma = \pi g^2 /m$ (equation (\ref{gamma1})) and $\Delta \chi$ is given by (\ref{Delta chi}). The interference term for the detector at $\xi' > \xi$  is given by
\beqn
\label{Jint1}
\begin{split}
\Jint = &2 \pi g \langle \dot{q}^{\dagger} \dot{\phi_{\rm out}}+ \dot{q} \dot{\phi}_{\rm out}^{\dagger} \rangle\\
=& -\gamma\sum_ke_k^2 \int_{-\infty}^\infty\int_{-\infty}^\infty
k'k''e^{i(k''-k')\tret}\alpha_k(k')\alpha_k^*(k'')  [i\Delta \chi^*(k'') + i\Delta \chi(k')] \ dk'\ dk''.
\end{split}
\eeqn

We can write the redshift in terms of proper time along the world line of the oscillator.  We have
\beqn
  \bar{\tau} \equiv \tau + \tau_s = \tau_s\left(1- \frac{1}{f^3} \right) = \frac{\tau_s}{f^3}(f-1)(f^2+f+1).
\eeqn
So the redshift factor in the flux is $\propto \bar{\tau}^2,$ the proper time measured to the horizon and so for the oscillator close to the horizon and the detector at infinity ($x \sim 1$, $x' \sim 0$), 
\beqn
\label{redshifted}
F = \left(\frac{\bar{\tau}}{9\tau_s}\right)^2(\Jint + \Jdir).
\eeqn

We now proceed to compare the direct flux and the interference term. In appendix 2 we look at the stationary phase approximation to the $k'$ and $k''$ integrals in equations (\ref{Jdir1}) and (\ref{Jint1}).  The result is that the impedance terms, $\Delta\chi$, are evaluated at the stationary points. This allows us to write $\Jdir$ as
\beqn
\begin{split}
\Jdir =  & \gamma^2 \int_{0}^{\infty} \frac{dk}{k}  \int_{-\infty}^{\infty} dk''  \int_{-\infty}^{\infty}\ dk' \Big\{k' k''  e^{i(k''-k') \pt_{\rm ret}} \\
&  \times \left[\alpha_k^*(k'')\alpha_k(k')|\Delta \chi (k_{\alpha})|^2 +\beta_{k}^*(k'')\beta_{k}(k')  |\Delta \chi (k_{\beta})|^2\right]\Big\}.
\end{split}
\label{dir-noise}
\eeqn
where we have inserted the stationary points 
\beqn
k'(k) = k_{\alpha} = \frac{(-3k \pt_{\rm ret})}{\bar{\tau}}\ \  \mathrm{and}\ \ k'(k) = k_{\beta} = -2k 
\eeqn
given in appendix 2.  We now compare this with the interference term. Again, we use the stationary phase approximation to justify writing $\Jint$ (equation (\ref{Jint1})) as
\beqn
\Jint=-2\gamma \int_{0}^{\infty} \frac{dk}{k} \int_{-\infty}^\infty\int_{-\infty}^\infty
e^{i(k''-k')\tret}\alpha_k(k')\alpha_k^*(k'')[i\Delta \chi^*(k_{\alpha}) + i\Delta \chi(k_{\alpha})].
\eeqn
We now use the fluctuation-dissipation theorem (appendix 4) to write this as
\beqn
\Jint = -2 \gamma^2 \int_{0}^{\infty} \frac{dk}{k} \int_{-\infty}^\infty\int_{-\infty}^\infty
e^{i(k''-k')\tret}\alpha_k(k')\alpha_k^*(k'')|\Delta \chi(k_{\alpha})|^2.
\eeqn
Comparing with equation (\ref{dir-noise}) we see that the total flux (up to the redshift factor) is
\beqn
\label{totnoise}
\begin{split}
\Jdir+\Jint =& \gamma^2 \int_{0}^{\infty} \frac{dk}{k}  \int_{-\infty}^{\infty} dk''  \int_{-\infty}^{\infty}\ dk' \Big\{ k' k'' e^{i(k''-k') \pt_{\rm ret}} \\
& \left[-\alpha_k^*(k'')\alpha_k(k')|\Delta \chi (k_{\alpha})|^2+\beta_{k}^*(k'')\beta_{k}(k') \right]| \Delta \chi (k_{\beta})|^2 \Big\}\\
\equiv & - \mathcal F_\alpha + \mathcal F_\beta.
\end{split}
\eeqn
Note that if it were the case that $\alpha_k(k') = \beta_k(k')$, then $k_{\alpha} = k_{\beta}$ and this expression vanishes. This accords with our result for the case of a constantly accelerated oscillator in flat spacetime. 
\\

In the case that  $\alpha_k(k') \ne \beta_k(k')$, the flux at the detector does not vanish. If we place the detector closer to the hole than the oscillator, the interference term now involves the ingoing modes. Thus it makes a contribution $-2\mathcal F_\beta$ to the flux which therefore becomes $-F$. Thus, if the energy radiated to infinity by the infalling oscillator is positive, the energy radiated into the hole is negative.   
\\

Given our particular forms for $\alpha_k(k')$ and $\beta_k(k')$ corresponding to a freely-falling oscillator we proceed to show that the flux has a blackbody spectrum (modified by the oscillator impedance)  as the oscillator approaches the black hole. 
\\

We can write the fluxes in equation (\ref{totnoise})  as
\beqn
\label{Ialphabeta}
\mathcal F _\alpha = \gamma^2 \int_{0}^{\infty} \frac{dk}{k} I_{\alpha}(k)I^{*}_{\alpha}(k)
\eeqn
where 
\beqn
\label{J_1}
I_{\alpha}(k) = \int_{-\infty}^{\infty}k' \alpha_k(k')e^{-ik'\tau} \Delta \chi(k')dk'
\eeqn
with corresponding definitions for $\mathcal F_\beta$ and $I_\beta(k)$.  The susceptibility factor is peaked around $k' = \pm \oc$.  The stationary phase approximation to the integral for $k>0$ will turn out to require $k' <0.$ Thus we can make the replacement $\Delta \chi(k') = \chi^* (k') - \chi(-k') \approx \chi^*(k').$ This corresponds to keeping the energy conserving terms $a^{\dagger} b + a b^ {\dagger}$ in the Hamiltonian. We shall find that the remaining terms that give rise to the Unruh effect (coming from $a^{\dagger}b^{\dagger} + ab$  in $\mathscr{H}$) give a contribution $\gamma/\oc$ smaller.    
\\

We now have
\beqn
I_{\alpha}(k) = \frac{2Mk}{\pi} e^{2\pi Mk} e^{-2 \pi iMk}e^{i (\pi/2 - 2\oc /\gamma)}(4M)^{4iMk} \Gamma(-4iMk)J_1
 \eeqn
where 
\beqn
\label{J_1D}
J_1 = \int_{-\infty}^{\infty}dk'e^{-ik'\bar{\tau}}(3k'+k)^{4iMk}f(k')
\eeqn
with
\beqn
f(k') = k'(3k+k')^{-1}\Delta \chi(k').
\eeqn
Evaluating $J_1$ for large $Mk$ by stationary phase (appendix 2) gives
\beqn
J_1 \sim (2\pi)^{1/2} e^{-4iMk +3ik\tau - i \pi /4} (4Mk)^{4iMk-1/2}\bar{\tau}^{-4iMk} f\left(\frac{-3k\tau}{\bar{\tau}}\right).
\eeqn
The contribution to the flux from $I_{\alpha}(k)I^{*}_{\alpha}(k) $ is therefore 
\beqn
\label{alphaflux}
\begin{split}
-\mathcal F_\alpha = &- 9\gamma^2 \left(\frac{\tau}{\bar{\tau}}\right)^2 \int_0^{\infty}kB(-8\pi Mk) \frac{dk}{(\oc + k'(k))^2 + \gamma^2 /4}\\
=& - \gamma^2 \int_{0}^{-\infty}k'B\left(2\pi \bar{\tau}k'\right)\frac{dk'}{(\oc+k')^2 + \gamma^2/4},
\end{split}
\eeqn
where we have substituted $k'(k) = -3k\tau / \bar{\tau}$ and taken the limit $\tau \rightarrow \tau_s,$ and where the black-body function $B(x)$ is:
\beqn
B(x)=\frac{1}{e^x-1}=-B(-x)-1.
\eeqn 

Finally, we can extend the integration to the full range with an error of order $\gamma / \oc$ since
\begin{equation*}
\begin{split}
\int_{0}^{\infty} \frac{dk}{(\oc+k)^2 + \gamma^2/4}=&\int_{\oc}^{\infty}\frac{dx}{x^2 + \gamma^2/4} = \frac{1}{\gamma}\int_{\oc/\gamma}^{\infty}\frac{dy}{y^2+1}\\
\sim &\frac{1}{\gamma} \left(\frac{\gamma}{\oc}\right).
\end{split}
\end{equation*}
This contribution to the flux is therefore 
\beqn
-\mathcal F_\alpha = -\gamma^2 \int_{-\infty}^{\infty}k'B\left(2\pi \bar{\tau}k'\right)\frac{dk'}{(\oc+k')^2 + \gamma^2/4}
\eeqn
where the integration is now over the full range of $k'$.
\\

Two comments are required here. The appearance of the blackbody factor arises from the Fourier analysis  of the time dependence of the oscillator trajectory. It may seem strange that approximating the Fourier integral in (\ref{alpha}) and its inverse transform in (\ref{J_1}) introduces a blackbody factor. This arises through our treatment of the gamma functions which for consistency should strictly be evaluated in the asymptotic (large $Mk$) limit. Doing this would lead to the Wien tail of the blackbody emission. However, we shall stick with precedent (dating back to Hawking's original paper) and retain the full blackbody form.  
\\

The second comment is the equally apparently strange way in which taking the Fourier transform followed by its inverse leads us merely to a form for $k'(k)$ in the oscillator susceptibility which is not contributing to the phase. In fact, including the phase of the susceptibility in the stationary phase approximation makes no difference to the result to the lowest order in $\gamma/\oc$. An alternative approach is to note that the susceptibility  is peaked around $k'=\oc$ and to expand the integrand about this point. To the accuracy of our approximation this leads to the same final result.      
\\

To calculate the contribution from the ingoing ($\beta$) modes we start from  (\ref{Ialphabeta}) with
\beqn
I_{\beta} =4 \left(\frac{2M}{\pi k}\right)^{1/2} e^{iMk(-10/3 + 4 \ln 2) + i\pi /4} \int_{-\infty}^{\infty}k' \Delta \chi (k') e^{i[-k'\bar{\tau}+(2M/k)(k+2k')^2]}dk'.
\eeqn
We evaluate this again  by stationary phase.  We let 
\beqn
\phi(k') = -k'\bar{\tau} + \frac{2M}{k} (2k'+k)^{2}.
\eeqn
The stationary point is
\beqn
k'(k)=-\frac{k}{2}\left(\frac{\bar{\tau}}{8M}+1 \right)
\eeqn
from which we obtain
\beqn
\int_{-\infty}^{\infty}k' \Delta \chi (k') e^{i[-k'\bar{\tau}+(2M/k)(k+2k')^2]}dk' \sim -\frac{k^{3/2}}{2} \left(\frac{2\pi}{M}\right)^{1/2}\left(\frac{ \bar{\tau}}{8M}+1\right) e^{i[\bar{\tau}k/2 - \bar{\tau}^2 k/(32M)]}\Delta\chi(k'(k)).
\eeqn
The dominant term in $\Delta \chi$ comes from $\chi(+k')$.  Thus, to lowest order in $\gamma$, as $\bar{\tau} \rightarrow 0$ 
\beqn
\begin{split}
\mathcal F_\beta=\gamma^2 \int_{0}^{\infty}\frac{dk}{k} I_{\beta}I^{*}_{\beta} & \sim  \gamma^2 \int_{0}^{\infty}\frac{k}{16}|\chi(k'(k))|^2 dk \\
& \sim -\gamma^2  \int_{-\infty}^{\infty} \frac{k' dk'}{(\oc + k')^2 + \gamma^2 /4}\\
\end{split}
\eeqn

Now use the relation $B(x) = -B(-x)-1$ to write (\ref{alphaflux}) as 
\beqn
-\mathcal F_\alpha = -\gamma^2\int_{-\infty}^{\infty}k'[-B(2\pi \bar{\tau}k')-1] \frac{dk}{(\oc + k'(k))^2 + \gamma^2 /4}.
\eeqn
This gives a positive frequency blackbody term plus the zero point energy that cancels the contribution from $I_{\beta}$. The total flux is therefore
\beqn
\label{dir+int}
\Jdir+\Jint=\gamma^2\int_{-\infty}^{\infty}k'[B(2\pi \bar{\tau}k')] \frac{dk}{(\oc + k'(k))^2 + \gamma^2 /4}.
\eeqn  
As promised, the result is a blackbody spectrum modulated by the oscillator susceptibility. 
\\

We now proceed to evaluate the remaining  integral over wave number in (\ref{dir+int}). This has the form of a smoothly varying factor multiplied by the susceptibility which (for an under-damped oscillator) is peaked around $k'=-\oc$. (We have $k'(k) = -3k\tau/\bar{\tau} < 0 $ since $k>0$, which justifies the inclusion of only the terms in $k' + \oc$ in (\ref{J_1})). We have
\beqn
\int_{0}^{-\infty} \frac{k'dk'}{(\oc + k')^2 + \gamma^2/4} = \int_{-\infty}^{\infty}\frac{dx}{x^2 + \gamma^2/4} + \mathcal O (\gamma/\oc) = \frac{2 \pi} { \gamma }
\eeqn
giving
\beqn 
\label{F-flux}
F\sim 2 \pi \left(\frac{\bar{\tau}} {9 \tau_s} \right)^2 \gamma \oc B(2\pi \oc (-\bar{\tau}))
\eeqn
for $ |\bar{\tau}| > \oc ^{-1}$ and $|\bar{\tau}| < \sim M$, where we have re-instated the redshift factor (equation (\ref{redshifted})).The blackbody factor peaks at $\bar{\tau}\oc  \sim 1$ and the flux at the peak is of order $ \gamma \oc / (M^2 \oc^2)$ or $ (t_d/ t_{in})(\lambda/R_s)$ where $t_d \sim 1/\gamma$ is the decay (or equilibration) time, $t_{in} \sim M$ is the infall time, $\lambda = 1/\oc$ is the wavelength of the oscillator and $R_s$ is the radius of the black hole. 
\\

For $\bar{\tau} < \oc^{-1}$ the oscillator susceptibility is no longer peaked, but we can evaluate the flux as follows. For   $\oc \bar{\tau} \rightarrow 0$ we have
\beqn
|\chi(k')|^2 = [(\oc + k'(k))^2 + \gamma^2 /4]^{-1} \sim \bar{\tau}^2 (9k^2 \tau^2 +\bar{\tau}^2 \gamma^2 /4)^{-1}.
\eeqn
The contribution to the total flux is
\beqn
\begin{split}
F_{0} &= 9\gamma^2\left(\frac{\tau}{\bar{\tau}}\right)^{2}\int_{0}^{\infty}k B(8\pi Mk) \frac{\bar{\tau}^2dk}{9k^2 \tau^2+\bar{\tau}^2 \gamma^2 /4}\\
&\sim \gamma^2\int_{0}^{\infty}  B(8\pi Mk) \frac{d(MK)}{Mk}.
\end{split}
\eeqn
as $\bar{\tau} \rightarrow 0$ and $\tau \rightarrow \tau_s$. The divergence at the lower limit can be dealt with by insisting that we are considering the case $Mk > 1$ or by a more accurate treatment of the stationary phase, which brings in an extra factor of $4Mk(16M^2 k^2 +1)^{-1}$ (see appendix 3). In either case the contribution is proportional to $\gamma^2$ which is $\gamma /\oc$ smaller than $F$. We can therefore ignore this contribution.  
 \\

We can estimate the total energy, $E$,  emitted by integrating (\ref{F-flux}) over time, $t'$, at the detector. Taking into account the redshift factors from (\ref{Fdir}), and $dt' = \frac{dt'}{d\tau}d\tau = \frac{\tau_s}{\bar{\tau}}d\bar{\tau}$, we have
\beqn
\label{E-B}
\begin{split}
E=&\pi \gamma \oc \int_{-\infty}^{\oc^{-1}} B(-2\pi\oc \bar{\tau}) \left(\frac{\bar{\tau}}{\tau_s}\right) d\bar{\tau}\\
=& \frac{3\gamma}{16 \pi M \oc} \int_{2\pi}^{\infty}xB(x) dx.
\end{split}
\eeqn
We can write this in terms of the infall time $t_{\rm in}$, the decay time of the oscillator $t_{\rm d}$ and the  wavelength $\lambda = 1/\oc $ as 
\beqn
E\sim \oc \frac{\gamma M}{\oc^2 M^2} \sim \oc \left(\frac{t_{\rm in}}{t_{\rm d}}\right)\left(\frac{\lambda}{R_s}\right)^2
\eeqn
for $\lambda <\sim R_s$. Note how the time-dependence of the infall in (\ref{E-B}) spreads the expectation value of the energy from the oscillator at frequency $\oc$ (or more precisely, the renormalised frequency) into a blackbody spectrum.

   %%%%        DISCUSSION    %%%%%%%%%

\section{Discussion} 

We can summarise our conclusions as follows. 
The infalling oscillator emits positive energy to infinity and negative energy into the black hole. This arises from the difference between the outgoing and ingoing modes (brought about by the cross-term in the metric, which provides the time asymmetry).  The flux comes from a distance $ > 1/\oc$ from the horizon. This suggests that as long as the ingoing modes are close to Painlev\'{e}-Gullstrand modes the oscillator radiates even if the true horizon does not form \cite{bardeen2014}.  
\\

The dominant effect comes from the energy conserving term in the Hamiltonian. (The atom is de-excited and emits a (scalar) photon. This is possible because the usual balance between excitation and de-excitation that results in the stability of  the ground state is disturbed by the difference between ingoing and outgoing modes. Thus the ground state is no longer stable moment by moment. The Unruh terms in the Hamiltonian in this case yield a blackbody flux to infinity (and a negative flux into the hole), which is a factor $\gamma/\oc $ smaller than the dominant terms. In this model, the Unruh effect would dominate by neglecting the back-reaction of the field on the oscillator. Indeed, if we replace $\chi(k')$ in (\ref{Jint1}) by $\lim_{\gamma \rightarrow 0} \chi(k') = i \pi \delta (k'+\oc)$ and put $\nu =- 2k/3$ (the minus sign allowing for the energy non-conserving term in $\mathscr{H}$) we obtain the expression for the energy flux equivalent to that in Scully et al.\cite{scully2018quantum} (although in the Boulware vacuum of their choice, this would be cancelled by the contribution from the direct flux).   
\\

As a simple model we can imagine a shell of oscillators (or atoms) collapsing to form a black hole. They radiate a blackbody flux to infinity and a negative energy flux into the (putative) black hole. But the emission from the oscillators is not the Hawking radiation: the flux from the oscillators depends on the strength of the matter-field coupling, so is not independent of their material properties; it occurs on a collapse timescale, not the Hawking timescale; and furthermore the flux is independent of the mass of the black hole (although spectrum depends on $M$ and the total energy radiated depends on $M$ through the infall time).  Nevertheless, the model may provide some useful hints. 
\\

In a self-consistent picture, the black hole is bathed in both incoming negative energy fluctuations and outflowing positive energy. The fluctuations in this radiation field will perturb the hole and cause it to radiate. Thus, in the fuller picture, the emission can be seen as a two-quantum process. Furthermore, the second quantum carries information about the first. In other words, it is conceivable that information is not lost in the process in much the same way that it is not lost in the "burning paper" illustration. In other words, we need to consider the reaction of the hole not just to its own radiation, but to that from the infalling matter (which, if nothing else, will be coupled to gravity). This may alter the argument from timescales \cite{mathur2011}. 
\\

One final speculation based on the  model presented here. The Lamb shift is usually absorbed into the mass of the oscillator by renormalisation and therefore in effect neglected. In a Bohr atom $\oc \propto m$, the electron mass, so the energy radiated (which we found to be proportional to $\oc$) is the (renormalised) mass. This suggests we need to incorporate a theoretical account of the origin of mass into the theory if we are to understand the quantum mechanics of black holes.

%%%% APPENDIX

%**********FOURIER TRANSFORMS************

\newpage
\section*{Appendix 1: Fourier Transforms $\alpha_k(k')$ and $\beta_k(k')$ of the Modes}

We have defined $\alpha_k(k')$ as the Fourier transform of the out-going modes evaluated along the worldline of the
oscillator, 
\beqn
\alpha_k(k')=\fr{1}{2\pi}\int_{-\infty}^{-\pt_s} e^{ik(\pt-\xi(\pt))}e^{ik'\pt}\ d\pt, 
\label{alpha1}
\eeqn
where $\tau = - \pt_s = - 4M/3$ is the location of the event horizon.  The worldline of the oscillator in free-fall is, in 
Painlev\'{e}-Gullstrand coordinates, 
\beqn
r_s(\pt)=\left(\fr{9M}{2} \right)^{\fr 1 3}(-\pt)^{2/3}.
\eeqn
with the function $\xi(r)$ defined as:
\beqn
\xi(r)=r+2\sqrt{2Mr}+4M \ln\left(\sqrt{\fr{r}{2M}}-1 \right).
\eeqn
When evaluated on the worldline of the oscillator this becomes:
\beqn
\xi(\pt)=\left(\fr{9M}{2} \right)^{1/3}(-\pt)^{2/3}+(48M^2)^{1/3}(-\pt)^{1/3}+4M\ln\left[
\left(\fr{3}{4M}\right)^{1/3}(-\pt)^{1/3}-1\right]
\label{xi(t)}
\eeqn
The behaviour of the integral is dominated by the argument of the exponential at the endpoint of the range, namely near the horizon. We therefore expand $\xi(\tau)$ about the horizon with
\beqn
\pt_0=-\pt_s-\eps
\eeqn
where $\eps>0$ is small compared to $\pt_s$.  Note that this expansion means that we remain in the exterior region.  Performing the expansion for each of the terms in (\ref{xi(t)}) yields the asymptotic form
\beqn
\xi(\eps)\approx 6M+2\eps+4M\ln(\eps)-4M\ln(4M),
\eeqn
and
\beqn
\pt_0-\xi(\eps)\approx -\fr{22M}{3}-3\eps-4M\ln(\eps)+4M\ln(4M).
\eeqn
Using this expansion in (\ref{alpha1}) means that
\beqn
\alpha_k(k')=\fr{1}{2\pi}e^{-iM(22k+4k')/3}(4M)^{4iMk}\int_0^\infty e^{-i(3k+k')\eps}\eps^{-4iMk}\ d\eps.
\eeqn

The integral over $\eps$ may be converted into a Gamma function.  To do this, we use the result from \cite{inttransvol1}: 
\beqn
\Gamma(z)=s^z\int_{0}^{\infty\ e^{i\delta}}e^{-st}t^{z-1}\ dt,
\eeqn
with $-(\pi/2+\delta)<\mrm{arg}\ s<\pi/2-\delta$, $\Re(z)>0$.  This result holds for $\mathrm{arg}\ s+\delta=\pm\pi/2$ provided $0<\Re(z)<1$.  Applying this to our $\eps$-integral we obtain, after some algebra, that for $(3k+k') > 0$\\
\beqn
\label{alphaplus}
\alpha_k(k')=\fr{2M}{\pi}ke^{-2\pi Mk}e^{-iM(22k+4k')/3}(4M)^{4iMk}(3k+k')^{4iMk-1}\Gamma(-4iMk)
\eeqn
and for $(3k+k') < 0$
\beqn
\label{alphaminus}
\alpha_k(k')=-\fr{2M}{\pi}ke^{2\pi Mk}e^{-iM(22k+4k')/3}(4M)^{4iMk}(-3k-k')^{4iMk-1}\Gamma(-4iMk).
\eeqn

We now determine an approximate expression for
the Fourier transform $\beta_k(k')$ of the in-going modes along the worldline of the
oscillator. 
\beqn
\begin{split}
\beta_k(k')=&\fr{1}{2\pi}\int_{-\infty}^{\pt_s} e^{ik(\pt+\eta(\pt))} e^{ik'\pt}\ d\pt \\
=&\fr{1}{2\pi}\int_{-\infty}^{\pt_s} e^{ikM\phi(\tau)} \ d\tau
\end{split}
\label{beta2}
\eeqn
with
\beqn
\eta(r)=r-2\sqrt{2Mr}+4M\ln\left(\sqrt{\fr{r}{2M}}+1\right).
\label{eta(r)}
\eeqn
We are treating $Mk$ as a large parameter in the integrand. Since $\phi(\tau)$ does not have a stationary point in the range (or in $(-\infty, 0)$), we expand about the end point on the horizon. (In fact, our result for the energy flux is independent of the point chosen to the accuracy of the approximation.)  
Define $t = -\tau/\tau_s = 3 \tau / 4 M$. Then, expanding $\phi(\tau)$ to second order gives
\beqn
k\phi(t) = -\frac{10}{3} + 4 \ln 2 -\frac{1}{3} (t-1) + \frac{1}{18}(t-1)^2 + \frac{4}{3} k't. 
\eeqn  
In terms of  $y=\sqrt{k\tau_s/24} (t-1)$, the integral for $\beta_k(k')$ becomes
\beqn
\beta_k(k') \sim \frac{1}{\pi^2}\left( \frac{6\tau_s}{k}\right)^{1/2} \int_{-\infty}^{\infty} dy \exp \left\{ \mp i \left [y+ \sqrt{\frac{2M}{|k|}(k + 2k')}\right]^2 +\frac{2M}{k} (k+2k')^2 \right\}
\eeqn
with the signs $\pm$ according as $k > 0$ or $k<0$. We justify the extension of the range to $-\infty$ as follows. We are going to use this expression in the approximate evaluation of the energy flux by stationary phase about the stationary point $2k'+k = -ik(t-1)/3$, with $k>0$, which gives a term $-4y^2$ in the exponential. Thus the integrand converges rapidly as $y \rightarrow +\infty$. We obtain
\beqn
\beta_k(k')= 2 \sqrt{\frac{2 M}{\pi |k|}}e^{i Mk(-10/3 + 4 \ln(2)) \pm i\pi /4 + i\tau_s k' -i\frac{2M}{k}(k+2k')^2}.
\eeqn

\section*{Appendix 2: 
Evaluation of integrals by stationary phase}
\label{stat-phase}
We want to evaluate integrals of the form
\beqn
\label{stat-phase}
I(k) = \int_{-\infty}^{\infty} e^{-iu\bar{\tau}+4iMk\ln u}f(u) \frac{du}{u}
\eeqn
where $u=(3k+k')$ and $\bar{\tau} = \tau + \tau_s < 0$ and $f(u)$ is a smoothly varying function of $u$. We have 
\beqn
I(k) = \int_{-\infty}^{\infty}e^{i\phi(u)}f(u) du
\eeqn
where
\beqn
\phi(u) = -u\bar{\tau}+(4Mk+i)\ln u.
\eeqn
The stationary point occurs at $u=u_0=\frac{-4Mk+ i }{\bar{\tau}}$. We therefore have
\bdis
 \phi''(u_0)=-\bar{\tau}^2/(4Mk+i). 
\edis
Thus
\beqn
\begin{split}
I(k) &\sim  e^{i\phi(u_0)}f(u_0)\int_{-\infty}^{\infty}\exp \left[\frac{i}{2}\phi''(u_0)(u-u_o)^2\right]du\\
 &= e^{-i\pi /4}e^{i\phi(u_0)}f(u_0)\left[\frac{2\pi(4Mk+i)}{\bar{\tau}^2}\right]^{1/2}\\
 &= (2\pi)^{1/2}e^{-i\pi /4}e^{(1-4iMk)}\left(4Mk+i\right)^{4iMk-1/2}f((4Mk+i)/\bar{\tau})
\end{split}
\eeqn
and
\beqn
\begin{split}
I(k)I^{*}(k)&=2\pi e^2 (4Mk+i)^{4iMk}(4Mk-i)^{-4iMk} (16M^2 k^2 +1 )^{1/2}f\left(\frac{4Mk-i}{\bar{\tau}}\right)
f\left(\frac{4Mk+i}{\bar{\tau}}\right)\\
&=2\pi e^2 (16M^2 k^2 +1 )^{1/2}\exp[{-8Mk \tan^{-1} (4Mk)^{-1}}]\left|f\left(\frac{4Mk-i}{\bar{\tau}}\right)\right|^2.
\end{split}
\eeqn
In the large $Mk$ limit we have
\beqn
I(k) \sim  (2\pi)^{1/2}e^{-i\pi /4}e^{-4iMk}\left(4Mk\right)^{4iMk-1/2}f((4Mk)/\bar{\tau})
\eeqn
which is the form we would obtain directly from (\ref{stat-phase}) and which we use in the body of the text.

\section*{Appendix 3  Evaluation of the decay rate, $\gamma$} 
\label{gamma}
Let $\Gamma = \gamma /2 + i \Delta \omega$, then we have
\beqn
\begin{split}
\Gamma &=i \frac{\oc g^2}{2m} \int_{-\infty}^{\infty}\int_{-\infty}^{\infty}\int_{-\infty}^{\infty}\frac{dk}{k}\left[ \alpha^{*}_{k}(k'')\alpha_{k}(k')+\beta_k(k')\beta^{*}_k(k'')\right]\frac{e^{i(k''-k')\tau}}{\oc + k'} dk''dk'\\
&= \Gamma_{\alpha} + \Gamma_{\beta},
\end{split}
\eeqn
where $\alpha_k(k')$ and $\beta_k(k')$ are given by (\ref{Alpha1}) and (\ref{beta1}). We are going to evaluate the $k'$ and $k''$ integrals by stationary phase. It will turn out that the stationary point lies in the region $3k+k' <0$ so we use the corresponding form for the $\alpha_k(k')$. Then, from appendix 2, we have
\beqn
I_1=\int_{-\infty}^{\infty}\alpha_{k}^{*}(k'')e^{ik''\tau} dk'' = -\frac{2eMk}{(i \pi \bar{\tau})^{1/2}} e^{2\pi Mk} \Gamma(4iMk) e^{i \psi}\left(\frac{4Mk-i}{\bar{\tau}}\right)^{-4iMk-1/2}
\eeqn
where $\psi =4iMk + 22iMk/3-3ik\tau$. 
\\

Similarly
\beqn
 I_2 = \int_{-\infty}^{\infty}\alpha_{k}(k')e^{-ik'\tau} \frac{dk'}{\oc+k'}=I_{1}^{*}\times (\oc)^{-1}\left (1-\frac{3k\tau}{\oc \bar{\tau}}+\frac{i}{\oc \bar{\tau}}\right)^{-1}
\eeqn
In the large $Mk$ limit we have 
\beqn
(4kM \pm i)^{\pm 4iMk} = (16M^2 k^2 +1)^{1/2} \exp \left[ \pm 4Mk \tan^{-1} (\pm 1/4Mk)\right] \sim e^{-1}.
\eeqn

Finally we obtain
\beqn
i \frac{\oc g^2}{2m}I_1 I_2  = \frac{i \lambda^2 M}{\oc} \int_{-\infty}^{\infty} \frac{e^{4\pi Mk}}{\sinh (4\pi Mk)} (1+16 M^2k^2)^{-1/2} \left(1-\frac{3k\tau}{\oc \bar{\tau}} + \frac{i}{\oc \bar{\tau}} \right)^{-1} dk
\eeqn
To evaluate this we consider the large $Mk$ limit as $\tau \rightarrow \tau_s$. Provided the oscillator is more than a small fraction of $\tau_s$ outside he horizon, this implies that $\oc \bar{\tau}$ is large. (Large here means $>\ \mathcal O(1)$ since the integrand is exponentially decreasing as a function of $Mk$).  Thus we can write
\beqn
\frac{1}{ 1-\frac{3k\tau}{\oc \bar{\tau}} + \frac{i}{\oc \bar{\tau}}}=i\pi \delta\left(1-\frac{3k\tau}{\oc \bar{\tau}}\right) + {\rm PP} 
\eeqn
where PP stands for the principal part of the integral. The delta function then restricts $k$ to $k = \oc \bar{\tau}/3 \tau \sim - \oc \bar{\tau}/4M > 0$ and the contribution to $\gamma/2$ from the outgoing modes is  
\beqn
 \gamma /2 = -\frac{\pi \lambda^2}{2 \oc} B(-2\pi \oc \bar{\tau} \tau_s /\tau) = -\frac{\pi \lambda^2}{2 \oc} [-1 + B(2\pi \oc \bar{\tau})] \rightarrow \pi g^2 /4m
\eeqn
for $\bar{\tau}$ of order $M$ in the large $Mk$ limit (i.e.for $\bar{\tau} \gg \oc^{-1}$). 
\\

We now have to consider the ingoing modes (the terms in $\beta_k(k')$ in (\ref{beta1})). Note that the contribution to $\gamma$  comes from the zero point energy, so we expect the ingoing modes to make an equal contribution to $\gamma$. We want to evaluate
\beqn
\Gamma_{\beta} = i\lambda^2 \int_{-\infty}^{\infty}\int_{-\infty}^{\infty}\int_{-\infty}^{\infty}\frac{dk}{k}\beta_k(k')\beta^{*}_k(k'')\frac{e^{i(k''-k')\tau}}{\oc + k'} dk''dk'.
\eeqn
To ensure convergence of the Fourier integrals we add a small imaginary part to $k'$. Inserting the expressions for $\beta_k(k')$ from (\ref{beta1}) we find the condition for stationary phase is $k'(k)=-(k/2)(1+\bar{\tau} /M) \rightarrow -k/2$ and the contribution to $\gamma/2 $ is
\beqn
\frac{\gamma}{2}= i\lambda^2 \Re\left(\lim_{\epsilon \rightarrow 0}\left\{ \int \frac{dk}{k} \frac{1}{(\oc - 2k/3 - i \epsilon)}\right\} \right)= \frac{\pi g^2}{4m }.
\eeqn
(We ignore the infrared divergence from the pole at $k=0$ since we are restricted to large $Mk$; in any case, it is an artefact of 1+1 dimensions.)  
\\

Thus, in the vicinity of the black hole, the equal contributions from $\Gamma_{\alpha}$ and $\Gamma_{\beta}$ sum to give
\beqn
\frac{\gamma}{2} = \frac{\pi g^2}{2m}.
\eeqn 

%%%%%%%%          F-D Theorem    %%%%%%%%%%%%%

\section*{Appendix 4: The Fluctuation-Dissipation Theorem}

First we determine an expression for $|\chi^*(k)-\chi(-k)|^2$ using the definition given in (\ref{chi}):
\beqn
|\chi^*(k)-\chi(-k)|^2=\left|\fr{1}{\oc+k+i\gamma/2}-\fr{1}{\oc-k-i\gamma/2} \right|^2
\eeqn 
We have that:
\beqn
\Delta \chi(k)= \chi^*(k)-\chi(-k)=\frac{-2k-i\gamma}{(\oc + k +i\gamma/2)(\oc -k - i\gamma/2)}
\eeqn
Thus:
\beqn
|\chi^{*}(k)-\chi(-k)|^2=\fr{4k^2 +\gamma^2}{(\oc^2-k^2)^2 +\gamma^2 (\oc^2 + k^2)/2 + \gamma^4/4}
\label{mod chis}
\eeqn
We also have 
\beqn
i\chi(k)+i\chi^*(k)=-\fr{\gamma}{2(\oc+k)^2+\gamma^2/2}
\eeqn
and hence 
\beqn
\label{add chis}
(i\chi(k)+i\chi^*(k))-(i\chi(-k)+i\chi^*(-k))=\fr{4\gamma \oc k}{(\omega_c^2-k^2)^2 +\gamma^2 \oc k +\gamma^4/16}
\eeqn
The functions (\ref{mod chis}) and (\ref{add chis}) are peaked around $k=\omega.$ Also $\gamma \ll \oc.$ Thus
\beqn
 |\chi^*(k)-\chi(-k)|^2 \sim \fr{4\oc^2 }{(\oc^2-k^2)^2 +\gamma^2 \oc^2  }
\eeqn
and
\beqn
(i\chi(k)+i\chi^*(k))-(i\chi(-k)+i\chi^*(-k)) \sim\fr{4\gamma \oc^2}{(\omega_c^2-k^2)^2 +\gamma^2 \oc^2}
\eeqn
giving us the final result
\beqn
i\Delta \chi(k) + i \Delta \chi^{*}(k)=\gamma |\Delta \chi(k)|^2
\eeqn 
which is our fluctuation-dissipation theorem. Note that we use the theorem to establish the relationship between the direct and interference terms at the detector, but we use the exact expression for the impedances in evaluating the integrals over frequency.

%%%%%%%%%%%%%%%

\section*{Appendix 5: Derivation of the Langevin Equation}
\label{Louiselle}
The integro-differential equation for the oscillator annihilation operator $A(t)$ derived from the
Hamiltonian (\ref{hamiltonian}) with the rotating wave approximation is \cite{louisell73quantum} (where $\kappa_j  \rightarrow \lambda$ in our notation)
\beqn
\fr{dA}{dt}=-\sum_j |\kappa_j|^2\int_0^tA(t')e^{i(\omega_j-\oc)(t'-t)}\ dt'+G_A
\label{L1}
\eeqn
where
\beqn
G_A=-i\sum_j \kappa_jb_j(0)e^{-i\omega_jt}.
\label{G-L}
\eeqn
To ensure convergence of the Fourier transform we give $\omega_j$  a small imaginary part, $\omega_j \rightarrow \omega_j - i\epsilon$. 
\\

The Wigner-Weisskopff approach to solving this equation involves taking the Laplace transform of both sides, and then applying an approximation, which essentially allows the replacement of (\ref{L1}) by the Langevin equation  
\beqn
\label{Langevin2}
\fr{dA}{dt}=-\left(\fr 1 2 \gamma + i \Delta \omega\right)A(t)+G_A(t),
\eeqn  
where 
\beqn
\gamma=2\pi g(\oc)|\kappa(\oc)|^2 \ \ \mathrm{and} \ \ \Delta\omega =-\int\fr{g(\omega_j)|\kappa(\omega_j)|^2\
d\omega_j}{\omega_j-\oc}.
\label{L-gamma}
\eeqn
This approach works because the Laplace transform leads to an equation for $\tilde{A}(s)$, the Laplace transform of $A(t)$. This method is not available to us since the Laplace transform of (\ref{L1}) does not yield an equation for $\tilde{A}(s)$. (We could expand $\tilde{A}$ as a power series in $s$ but it is then difficult to control the approximation.) We  now show that the Langevin equation may be obtained using integration by parts. Returning to (\ref{L1}) we integrate by parts with respect to $t'$ :
\beqn
\int_0^t A(t')e^{i(\omega_j-\oc)(t'-t)}\ dt'=\fr{A(t)}{i(\omega_j-\oc)}-\frac{A(0)e^{-i(\omega_j - \oc)t}}{i(\omega_j - \oc)}-\int_0^t \fr{dA}{dt'} \fr{e^{i(\omega_j-\oc)(t'-t)}}{i(\omega_j-\oc)}\ dt'.
\label{IBP-L}
\eeqn
The integral in (\ref{IBP-L}) is of order $g$ smaller than the other terms, so can be neglected. The term in $A(0)$ represents the initial conditions and can be neglected in the differential equation. Thus (\ref{L1}) now becomes:
\beqn
\fr{dA}{dt}=iA(t)\sum_ j\fr{|\kappa(\omega_j)|^2
e^{i(\omega_j-\oc)t}}{\omega_j-\oc}+G_A.
\eeqn
We now convert the sum over $j$ into an integral over $\omega_j$:
\bdis
\sum_j|\kappa(\omega_j)|^2 \rightarrow \int_0^\infty g(\omega_j)|\kappa(\omega_j)|^2 \ d\omega_j,
\edis
and so, with $\omega_j\rightarrow \omega_j-i\epsilon $ then: 
\beqn
\fr{dA}{dt}=G_A+iA(t)\int_{0}^{\infty}\fr{|\kappa(\omega_j)|^2g(\omega_j)e^{i(\omega_j-\oc-i \epsilon)t}}{\omega_j-\oc-i\epsilon}\ d\omega_j
\label{L2}
\eeqn
Using the Sokhotski-Plemelj theorem \cite{inteqnsvol1}
\beqn
\begin{split}
\lim_{s\rightarrow 0}\int_{0}^{\infty}\fr{|\kappa(\omega_j)|^2g(\omega_j)e^{i(\omega_j-\oc-is)t}}{\omega_j-\oc-is}\ d\omega_j=&
\mathcal P \left(\int_{0}^{\infty}\fr{|\kappa(\omega_j)|^2g(\omega_j)e^{i(\omega_j-\oc)t}}{\omega_j-\oc}\ d\omega_j\right)\\
&+i\pi \int_0^\infty |\kappa(\omega_j)|^2g(\omega_j)e^{i(\omega_j-\oc)t}\delta(\omega_j-\oc)\ d\omega_j
\end{split}
\eeqn
Thus, after integrating out the delta function: 
\beqn
\fr{dA}{dt}=G_A-(\pi|\kappa(\oc)|^2g(\oc)+i\Delta \omega)A(t)
\eeqn
which gives (\ref{Langevin2}) with $\gamma$ and $\Delta \omega$ as defined in (\ref{L-gamma})

%%%%%%%%%%%

%%%%%%%%%%%%%%%%%%%%%%%%%%%%%%%
\newpage
\bibliography{research}
\bibliographystyle{acm}

\end{document}